\documentclass[useAMS,usenatbib,usegraphicx]{mn2e}

\usepackage{amssymb}

\title[Stellar haloes outshine disc truncations in low-inclined spirals]{Stellar haloes outshine disc truncations in low-inclined spirals}
\author[Mart\'in-Navarro et al.]{Ignacio Mart\'in-Navarro$^{1,2}$\thanks{E-mail:
imartin@iac.es (IMN); trujillo@iac.es (IT); jhk@iac.es (JHK)}, Ignacio Trujillo$^{1,2}$\footnotemark[1], Johan H.  Knapen$^{1,2}$\footnotemark[1],
\newauthor
Judit Bakos$^{1,2,3}$ and J\"{u}rgen Fliri$^{1,2}$\\
$^{1}$Instituto de Astrof\'isica de Canarias, E-38200 La Laguna, Tenerife, Spain\\
$^{2}$Departamento de Astrof\'isica, Universidad de La Laguna, E-38205 La Laguna, Tenerife, Spain\\
$^{3}$Konkoly Observatory, Research Centre for Astronomy and Earth Sciences Hungarian Academy of Sciences, Budapest, Hungary}

\begin{document}

\date{ }

\pagerange{\pageref{firstpage}--\pageref{lastpage}} \pubyear{2014}

\maketitle
\label{firstpage}
\begin{abstract}
The absence of stellar disc truncations in low-inclined spiral galaxies has been a matter of debate in the last decade. Disc truncations are often observed in highly inclined galaxies but no obvious detection of this feature has so far been made in face-on spirals. Here we show, using a simple exponential disc plus stellar halo model based on current observational constraints, that truncations in face-on projections occur at surface brightness levels comparable to the brightness of stellar haloes at the same radial distance. In this sense, stellar haloes outshine the galaxy disc at the expected position of the truncations, forcing their studies only in highly inclined (edge-on) orientations.
\end{abstract}


\section{Introduction} \label{sec:intro}
Since truncations in the surface brightness profiles of edge-on galaxies were first noticed by \citet{van79} and \citet{van81,van82}, it has been clear that the radial light distribution of spiral galaxies does not decline following an exponential law along the whole disc as proposed in the earliest photometric studies \citep{patt40,devau58,free70}. However, over the last decade, numerous works have revealed a dichotomy in the observational framework while comparing truncations in the surface brightness profiles of edge-on and face-on spirals. In edge-on galaxies, these truncations are usually found at radial distances of four to five times the exponential scalelength of the disc \citep[e.g.,][]{bart94,bott95,degri01,kre02,van07}, whereas face-on galaxies show downbending\footnote{In the literature, these features in low-inclined systems are often incorrectly referred to as truncations.} breaks closer to the galactic centre, typically at around two or three disc scalelengths \citep{erw05,pt06,erw08}. Moreover, the way in which the surface brightness profile changes depends on the orientation of the galaxy with respect to the line of sight: edge-on galaxies exhibit a sharp \textit{truncation} in the radial light distribution, as opposed to a smoother \textit{break} observed in face-on spirals.

The strong observational difference between edge-on and face-on galaxies is reflected in the proposed explanations for the formation of such breaks and truncations. Two main mechanisms have been proposed. On the one hand, \citet{van87,van88} suggested that truncations could be related to the maximum angular momentum of the proto-galactic cloud. In this scenario, it is expected that the truncation happens at a radius of four or five times the exponential scalelength, which is in good agreement with the edge-on observational results. On the other hand, a threshold in the star formation of the galaxy can also explain the presence of a break in the surface brightness profiles of spiral galaxies \citep[e.g.,][]{fall80,ken89,elm94,sch04,elm06}. This threshold in the star formation would lead to a change in the stellar population around the break radius, as both photometric \citep{azzo,bak08,bak12} and resolved stellar population \citep{jong07,rs12} studies of face-on galaxies appear to confirm. Numerical simulations tend to support the idea of the breaks being caused by a threshold in the star formation of the galaxy \citep{deb06,ros08,pat09,ms09}. In addition, these numerical simulations place the break at radial distances which fit the face-on measurements quite well, e.g., \citet{ros08} find the break radius at 2.6 $h_\mathrm{r}$ whereas the average break in \citet{pt06} occurs at 2.5 $h_\mathrm{r}$ ($h_\mathrm{r}$ is the exponential scalelength of the inner disc before the appearance of the break).

The access to deep and high-quality imaging from both the Sloan Sky Digital Survey (SDSS) Data Release 7 \citep{sdss} and the \textit{Spitzer} Survey of Stellar Structure in Galaxies (S$^4$G) \citep{s4g} has allowed to shed more light on the break-truncation dichotomy. Thanks to these surveys, \citet{mart12} and \citet{com12} have shown that both breaks and truncations can be observed simultaneously in edge-on galaxies. Consequently, there is growing consensus that truncations, usually observed in edge-on galaxies, and breaks, initially measured in face-on systems, are not the same phenomenon. However, an unavoidable question remains unanswered: Why are we able to detect both breaks and truncations in edge-on galaxies but only the former in face-on orientations?

To complete the observational picture of the radial light distribution of spiral galaxies it is necessary to take into account the stellar halo. Only a few photometric studies \citep[e.g.,][]{zib04,zife04,jab10,zack12,bak12,truba12,pet} have been able to reach surface brightnesses $\mu_\mathrm{r}\sim 30 ~ \mathrm{mag~arcsec}^{-2}$, deep enough to study the stellar halo properties. Although the formation and evolution of stellar haloes around spiral galaxies remains an open question, most observations point to old and moderately metal-poor stars as the main component of these stellar haloes, probably assembled in merger/accretion events at redshifts $z > 1$ \citep{zib04,truba12,bak12}.

Trying to reconcile all the previous observational results, we have built a 3D disc galaxy model using an \textit{exponential disc} plus \textit{stellar halo}. Our model comfortably fits the current observational constraints on the photometrical structure of spiral galaxies. We show how this simple model is able to explain why different disc orientations prevent the general detection of truncations in low-inclined galaxies. Basically, the expected surface brightness of the stellar disc at the position of the truncation in a face-on orientation is so faint that this feature is outshone by the brightness of the stellar halo at that radial distance.

In Section 2 we present a compilation of observational results of face-on and edge-on galaxies; in Section 3 we describe the model and finally, in Section 4 we derive the main conclusions of this work.

\section{Observational constraints to the model}\label{sec:obs}

To introduce the observational framework regarding breaks, truncations and stellar haloes in spiral galaxies, we compile  in this section the current observational constraints that describe these features.

\subsection{Breaks}

\citet{pt06} found in their sample of 85 low-inclined galaxies that downbending breaks appear, on average, at a radial distance of 2.5 $\pm$ 0.6 $h_\mathrm{in}$, where $h_\mathrm{in}$ is the exponential scalelength before the break radius. \citet{pet11} measured a quite similar value of 2.5 $\pm$ 0.2 $h_\mathrm{in}$ for the break radius among a sample of 12 nearby galaxies, while \citet{erw08} and \citet{leo11} found slightly lower values (1.8 $\pm$ 0.2 $h_\mathrm{in}$ and 2.1 $\pm$ 0.5 $h_\mathrm{in}$, respectively) for the break radius of early-type galaxy discs (with morphological types from S0 to SB). Breaks  also present a characteristic ratio between the exponential scalelength before and after the break radius ($h_\mathrm{in} / h_\mathrm{out}$). In this sense, low-inclined spiral galaxies show a value for $h_\mathrm{in} / h_\mathrm{out} \sim 2$, with values typically varying from $h_\mathrm{in} / h_\mathrm{out} = 2.6 \pm 0.3$ \citep{erw08} to $h_\mathrm{in} / h_\mathrm{out} = 1.7 \pm 0.1$ \citep{leo11}. This scatter lies 
within the measurements of \citet{pt06} who found a mildly peaked distribution ranging between $h_\mathrm{in} / h_\mathrm{out} = $ 1.3-3.6

\subsection{Truncations}

The first results of \citet{van82} placed the truncation radius at a radial distance of 4.2 $\pm$ 0.6 $h_\mathrm{in}$, significantly larger than the break radius. \citet{bart94} proposed a truncation radius of 3.7 $\pm$ 1.0 $h_\mathrm{in}$ which is compatible also with the results of \citet{poh00}, who found an averaged truncation radius of 2.9 $\pm$ 0.7 $h_\mathrm{in}$. As discussed in \S\ref{sec:intro}, truncations are principally detected in edge-on galaxies. Nevertheless, \citet{mart12} and \citet{com12} have shown that edge-on galaxies can simultaneously host breaks too. Thus, if no distinction is made between breaks and truncations while studying edge-on systems, the measurements can be confusing both features (breaks and truncations). This could be the reason for the large scatter in the values of the truncation radii found in the literature. For the ratio $h_\mathrm{in} / h_\mathrm{out}$ between the exponential scalelengths before and after the truncation radius, \citet{mart12} found a mean value of 1.9 $\pm$ 0.4, very similar to the observed change in the exponential scalelength around the break radius. This measurement is in agreement with \citet{com12} who did not find significant differences between the $h_\mathrm{in} / h_\mathrm{out}$ distributions for breaks and truncations.

\subsection{Stellar haloes} \label{sec:halo}

The current observational constraints on stellar haloes are not as strong as those on breaks and truncations because of the very low surface brightnesses where these haloes start to be visible, typically below $\mu \sim 28 ~ \mathrm{mag~arcsec}^{-2}$ in the R band. Stellar haloes are usually detected as an excess of light  with respect to the exponential decline of the discs in the outer parts of spiral galaxies ($R \gtrsim$ 15 kpc) and they are observed not only in images \citep[e.g.,][]{truba12,bak12,pet}, but in particular using star-counting techniques \citep[e.g.,][]{fer02,mo05,fer07,ib09,jab10}. The shape and the surface brightness of the stellar haloes are the same in edge-on and  face-on galaxies. For example, \citet{bak12} found them at around $R \sim$ 20 kpc, with a typical surface brightness in the SDSS $r'$-band of $\mu_{r'} \sim 28~ \mathrm{mag~arcsec}^{-2}$ in a sample of low-inclined disc galaxies. In the edge-on galaxy NGC~3957, \citet{jab10} measured the stellar halo above 20 kpc above the galactic plane, at a surface brightness of $\mu_R = 28.5~ \mathrm{mag~arcsec}^{-2}$. \citet{wu02} detected the stellar halo of the edge-on spiral NGC~4565 at 22 kpc measured along the minor axis, having a surface brightness of $\mu_{6660\mathrm{\AA{}}} = 27.5~ \mathrm{mag~arcsec}^{-2}$

\section{Model} \label{sec:model}

To explain the proposed model in a clear way, this section is split into two parts. In the first subsection we describe the mathematical formulation of the model, along with the required information regarding the vertical distribution of the disc and the dust component. In the second part, we use this same model to reproduce the observed characteristics of two galaxies with similar properties in the $r'$-band, one seen with a low inclination (UGC~00929) and the other one observed in an edge-on orientation (UGC~00507).
\subsection{Mathematical description}\label{sec:math}
The disc component of our model is parametrised as an exponential function in both the radial \citep{free70,van88,poh07} and the vertical direction \citep{wain89,grijs97,degri01}. Although the vertical exponential scalelength can vary with galactocentric distance \citep[see][]{dp97}, for simplicity, we assume it is constant along the whole disc \citep{van81,van82,shaw90}. In the radial direction, the model is characterised by two changes (a \textit{break} and a \textit{truncation}) in the exponential behaviour. Thus, the three-dimensional light distribution of our synthetic galaxy can be expressed in cylindrical coordinates $(R,z)$ as:
\begin{equation} \label{eq:disc}
L(R,z) = L_0 \ L(R) \, e^{-|z|/h_\mathrm{z}} \ ,
\end{equation}
where $L_0 $ is the luminosity at the centre of the disc and the radial light distribution is given by:
\begin{equation}\label{eq:rad}
L(R) = e^{-R/h_\mathrm{I}} \Pi_{0,r_\mathrm{B}} + \ e^{-R/h_\mathrm{B}} \Pi_{r_\mathrm{B},r_\mathrm{T}} + \ e^{-R/h_\mathrm{T}} \Pi_{r_\mathrm{T},\infty} \ .
\end{equation}

As mentioned before, Eq. (\ref{eq:rad}) shows two changes in the exponential behaviour: the first one (\textit{break}) is located at a radial distance $R = r_\mathrm{B}$, while the second change (\textit{truncation)} occurs at $R = r_\mathrm{T}$. The light distribution is parametrised by the vertical exponential scalelength ($h_\mathrm{z}$) and also by the radial exponential scalelength before the break radius ($h_\mathrm{I}$), after the break radius ($h_\mathrm{B}$) and after the truncation radius ($h_\mathrm{T}$). The boxcar function $\Pi_{a,b}$ is defined as:
\begin{displaymath}
   \Pi_{a,b} = \left\{
  \begin{array}{l l}
    C_{a,b} & \mathrm{if \ } a \leq R < b\\
    0 & \mathrm{otherwise} \ ,\\
  \end{array} \right. 
\end{displaymath}
where $C_{a,b}$ is constant, and $L(R)$ is continuous over the whole $R$-domain.

Although a certain level of clumpyness is expected in the dust distribution, we have only considered the diffuse component for the dust attenuation, parametrising the extinction with a continuous spatial exponential function \citep{xi97,xi98}. This exponential approach has properly described the dust distribution in edge-on galaxies \citep{xi99,mi01} and a very careful description can be found in \citet[$\S$2.1]{pop11}. The extinction of the dust disc can be written as:
\begin{equation} \label{eq:dust}
 \kappa^\mathrm{dust}(\lambda,R,z) = \kappa^\mathrm{dust}(\lambda,0,0) \, e^{-R/h^\mathrm{dust}_R} \, e^{-|z|/h^\mathrm{dust}_z}\ ,
\end{equation}
where $\kappa^\mathrm{dust}(\lambda,0,0)$ is the extinction at the centre of the disc and $h^\mathrm{dust}_R$ and $h^\mathrm{dust}_z$ are the dust scalelengths in the radial and vertical directions.

Finally, the surface brightness profile of the stellar halo is modelled using a S\'ersic \citeyearpar{ser68} law which has been widely used as a proxy for the projected  light distribution of this component in spiral galaxies \citep[e.g,][]{jong08,cou11,bak12}. We do not describe the 3D structure of the halo (as we have done before for the disc) but we only model its projected surface brightness distribution. For simplicity, we assume that the stellar halo is spherical. The projected light distribution of the stellar halo is described as follows:
\begin{equation} \label{eq:halo}
 I^\mathrm{halo}(R) = I^\mathrm{halo}_\mathrm{eff} \exp \left\lbrace -b_\mathrm{n} \left[\left(\frac{R}{R_\mathrm{eff}}\right)^{1/n} - 1 \right] \right\rbrace \ ,
\end{equation}
where $I_\mathrm{eff}$ is the intensity at the effective radius $R_\mathrm{eff}$, $n$ is the so-called S\'ersic index and $b_\mathrm{n}$ is a function of the S\'ersic index that, in good approximation, is given by  $b_\mathrm{n} \simeq 2 n - 0.324$  \citep[e.g.,][]{truji01}. To have a fair comparison with previous studies, we fixed the S\'ersic index of the stellar halo to n=1. This choice properly represents the shape of the stellar halo surface brightness profile \citep{irwin,ib07,tanaka,pet}.

Eqs. (\ref{eq:disc}) to (\ref{eq:halo}) completely define the three-dimensional light distribution in our model. The observed surface brightness profile is given by the integration along the line of sight (LOS) of Eq. (\ref{eq:disc}), attenuated by the dust component following Eq. (\ref{eq:dust}) plus the contribution of the stellar halo in Eq. (\ref{eq:halo}). Thus, the radial surface brightness profile of our synthetic galaxy, observed in an arbitrary orientation, can be written as:
\begin{equation}
 I^\mathrm{obs}(R) = I^\mathrm{halo}(R) + \int_\mathrm{LOS} L(R,z) e^{-\tau(\xi,\lambda)} \ \mathrm{d}\xi \ ,
\end{equation}
where the optical depth $\tau(\xi,\lambda)$ is given by:
\begin{equation}
\tau(\xi,\lambda) = \int_\mathrm{LOS} \kappa^\mathrm{dust}(\lambda,R,z) \ \mathrm{d}\xi \ .
\end{equation}

\subsection{Comparison with the data}
The above model can simultaneously reproduce the typical surface brightness profiles of low and highly inclined spiral galaxies. In order to do illustrate this, we have selected two galaxies, UGC~00929 (low-inclined) and UGC~00507 (highly-inclined), from the SDSS Stripe 82. The SDSS Stripe82 is a very deep photometric survey covering 275~deg$^2$ around the Celestial Equator and reaching $\sim2$~magnitudes deeper than the standard SDSS survey. These ultra-deep data are crucial to constrain the model since, as mentioned in \S\ref{sec:halo}, the stellar halo starts to dominate the surface brightness distribution at around $\mu_{r'} \sim 28 ~ \mathrm{mag~arcsec}^{-2}$. Table \ref{tab:data} summarises the basic information about these two objects, found in the HyperLeda database \citep{patu03}.

\begin{table}
\centering
\begin{center}\begin{tabular}{l c c c }
\hline
\hline
Galaxy & \multicolumn{1}{c}{$M_{\mathrm{abs}}$} & Morph. & \multicolumn{1}{c}{$incl$} \\
       & ($B$-mag) & type   &     (deg)     \\
\hline 
UGC~00507     &  -20.34   &   Scd    & 80.0 \\
UGC~00929     &  -20.60   &   SABc   & 24.7  \\
\hline 
\end{tabular}
\caption{Global parameters of the two galaxies analysed in this Letter (from the HyperLeda database). \label{tab:data}} 
\end{center}
\end{table}

The SDSS $r'$-band radial surface brightness profiles were calculated in the same way that \citet{bak12} and \citet{mart12} did for low and highly inclined spiral galaxies, respectively\footnote{An extensive explanation on the surface brightness profile calculation process can be found in these two papers.}: for UGC~00929 elliptical apertures were placed along the galaxy disc, while for the edge-on galaxy, UGC~00507, the profile was measured by placing a constant-width slit onto the galactic plane. The latter profile is noisier than the first one since, at the same radial distance, the number of pixels within a slit is much lower than within an elliptical aperture. This means that, by measuring the radial profile through a slit, we are limited to surface brightnesses above $\mu_{r'} \lesssim 27 ~ \mathrm{mag~arcsec}^{-2}$, where the stellar halo contribution to the total surface brightness profile of the galaxy just starts to be important. To address this problem and complete the edge-on galaxy model below $\mu_{r'} \sim 27 ~ \mathrm{mag~arcsec}^{-2}$, we measured the stellar halo of UGC~00507 by using circular apertures centred on the galaxy.

Apart from these radial surface brightness profiles, additional information about the vertical light distribution and about the dust was required to fully constrain our model. Typically, the vertical exponential scalelength is found to be $h_\mathrm{z} \sim 1/10 \ h_\mathrm{r}$ \citep{van82,grijs96}. For the dust distribution we followed \citet{xi99}, assuming a characteristic vertical scalelength $h^\mathrm{dust}_z\sim 1/20 \ h_\mathrm{r}$ and a radial scalelength $h^\mathrm{dust}_R\sim 2 \ h_\mathrm{r}$.

Once the surface brightness profiles for the two galaxies were obtained, we built up two synthetic galaxies that reproduce both the face-on and the edge-on surface brightness distributions. The structural parameter set that defines our synthetic galaxy models is listed in Table \ref{tab:model}.

\begin{table*}
\centering
\begin{center}\begin{tabular}{l c l l c l l c l}
\hline
\hline
\multicolumn{3}{c}{Face-on case} & \multicolumn{3}{c}{Edge-on case}   \\
\multicolumn{3}{c}{(UGC~00929)}    & \multicolumn{3}{c}{(UGC~00507)}     \\
\hline 
$\mu^\mathrm{break}_{r'}$ & 24.7 & [$\mathrm{mag~arcsec}^{-2}$]    & $\mu^\mathrm{break}_{r'}$  & 24.0 & [$\mathrm{mag~arcsec}^{-2}$]    \\
$r_\mathrm{B}$            & 4.1  & [$h_\mathrm{I}$]                &  $r_\mathrm{B}$            & 3.0  & [$h_\mathrm{I}$]         \\
$h_\mathrm{B}$            & 1.7  & [$h_\mathrm{I}$]                &  $h_\mathrm{B}$            & 2.1  & [$h_\mathrm{I}$]      \\
$r_\mathrm{T}$            & 6.0  & [$h_\mathrm{I}$]                &  $r_\mathrm{T}$            & 4.5  & [$h_\mathrm{I}$]        \\
$h_\mathrm{T}$            & 3.4  & [$h_\mathrm{I}$]                &  $h_\mathrm{T}$            & 3.6  & [$h_\mathrm{I}$]         \\
\hline 
\end{tabular}
\caption{Modelled structural parameters of the galaxies used in this paper.\label{tab:model}} 
\end{center}
\end{table*}

\begin{figure*}
\begin{center}
\includegraphics[width=250pt]{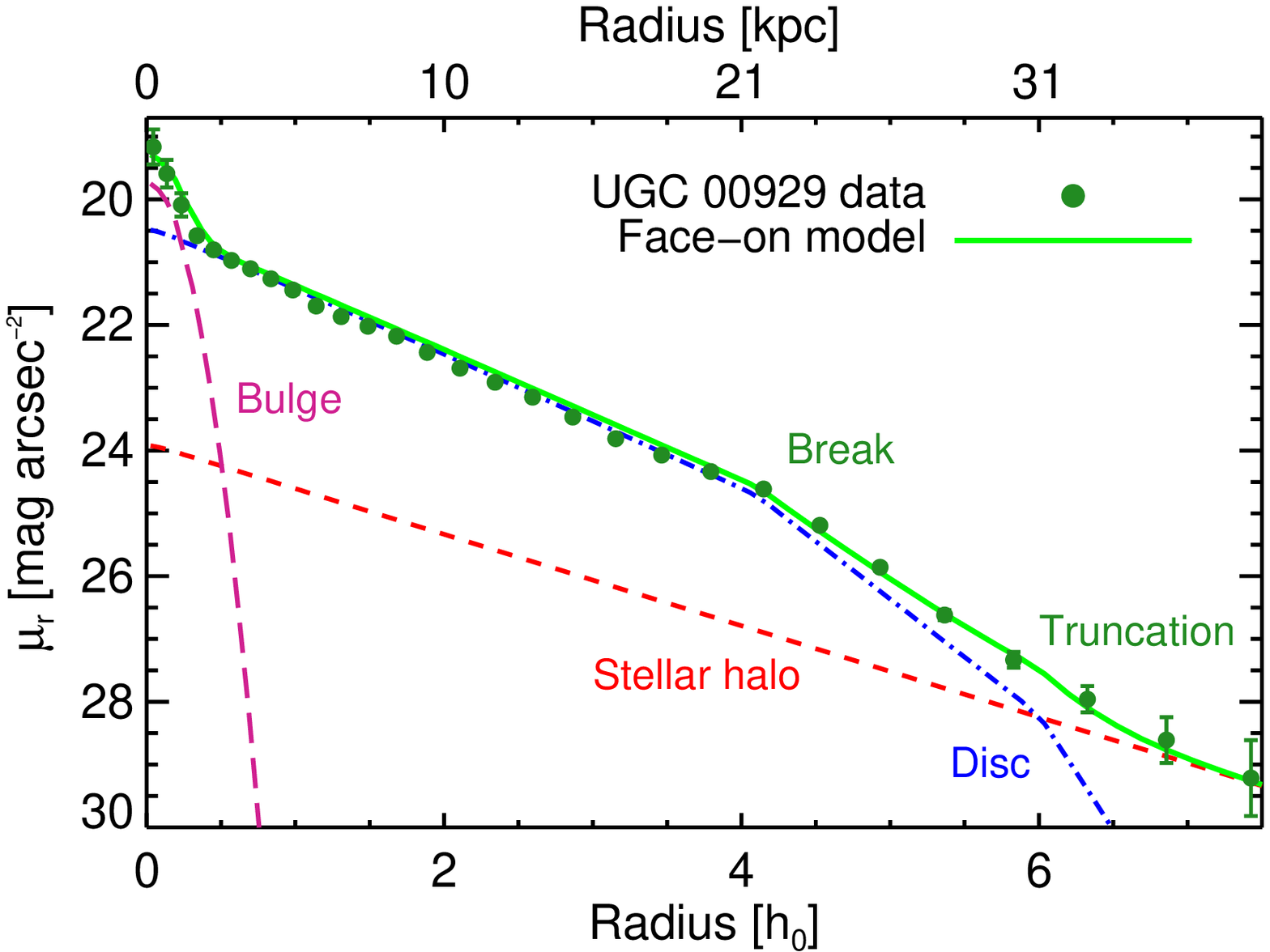}
\includegraphics[width=250pt]{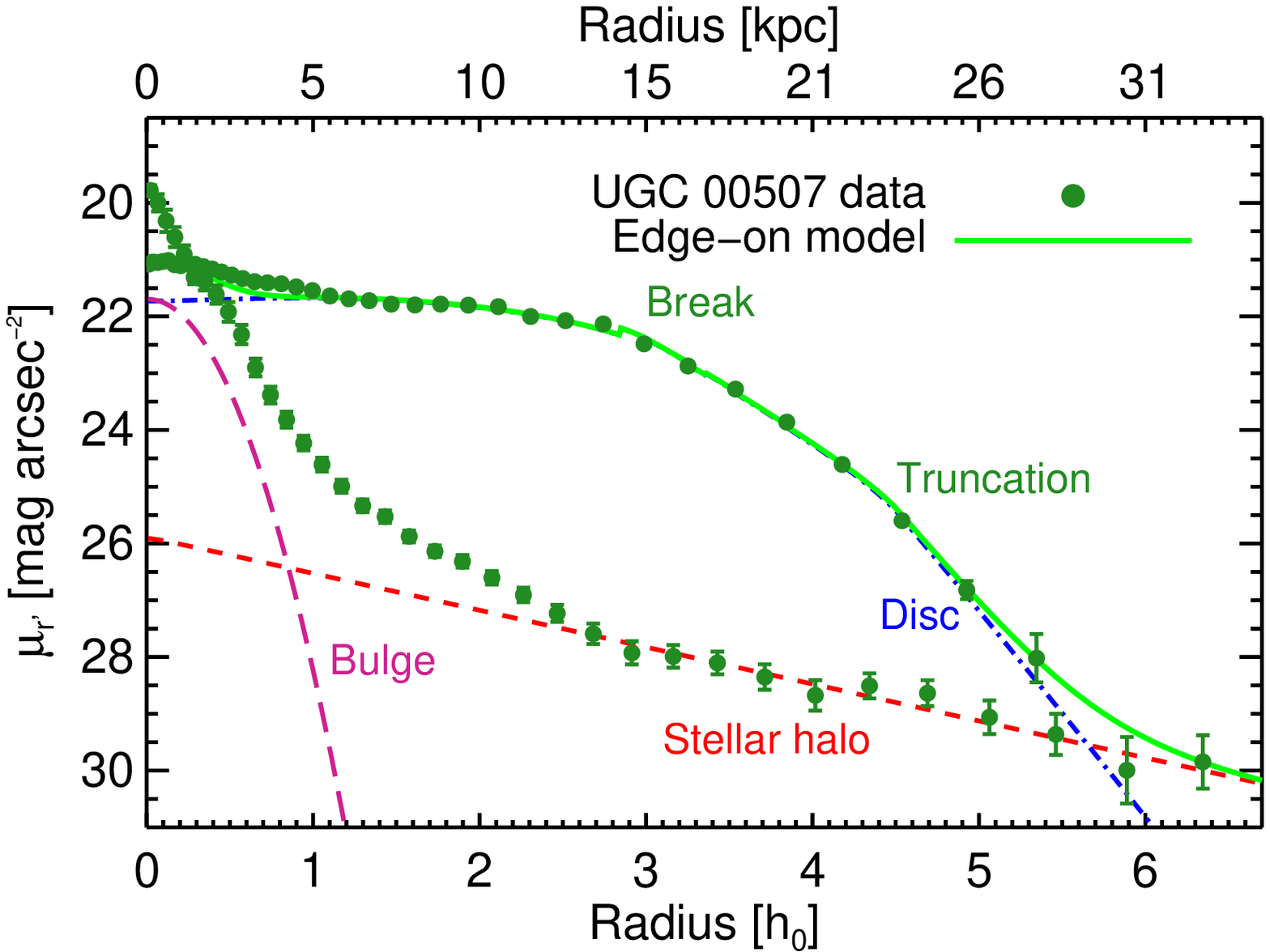} 
\end{center}
\caption{Surface brightness distributions of a face-on (UGC~00929) and an edge-on (UGC~00507) spiral galaxy. Our model fits to the surface brightness distribution are overplotted (green solid lines). We also show the different contributions to the final surface brightness profiles: blue dashed line for the disc, red dashed line for the stellar halo and yellow dashed line for the bulge. Because of the LOS integration, the observed radial surface brightness profile of the edge-on galaxy is brighter, allowing us to measure the truncation in the radial light distribution of the stellar disk above the stellar halo contribution. However, the halo brightness outshines the truncation in the face-on case since the surface brightness of the stellar disk and the halo are similar at the radial distance where the truncation happens. Note that for UGC~00507, the stellar halo radial profile was obtained using circular apertures. This is a fair approach under the assumptions made in our model (see \S\ref{sec:math}). Note, however, that the halo in UGC~00507 is $\sim$ 1 mag dimmer than in the face-on galaxy UGC~00929 at $\sim$ 30 kpc. This means that either the halo is intrinsically dimmer or that it does not obey a perfectly spherical symmetry.}
\label{img:break}
\end{figure*}

In Fig.\ref{img:break} we show how our simple model is able to reproduce the surface brightness profiles of low and highly-inclined galaxies. The structural parameters listed in Table \ref{tab:model} were not obtained by a systematised fitting process of the data showed in Fig.\ref{img:break} (dark green filled circles) since providing a precise characterisation of the observed data is not in the scope of the present Letter. Following a simplified approach, we tuned the free parameters in our model (see \S\ref{sec:math}) to obtain a reasonable representation of the data, taking into account the observational constraints detailed in \S\ref{sec:obs}. That is the reason why no uncertainties are given for the structural parameters showed on Table 2.

\section{Discussion and Conclusions}

\begin{figure}
\begin{center}
\includegraphics[width=240pt]{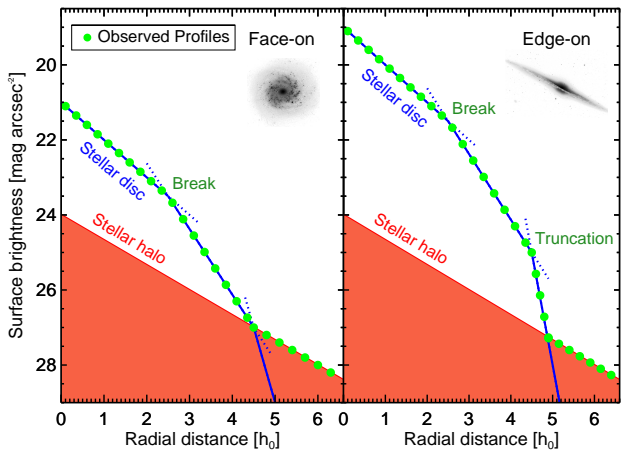}
\end{center}
\caption{Schematic representation of the observed surface brightness profiles of a face-on and an edge-on disc galaxy according to our interpretation. The surface brightness profile of the stellar hallo component is the same in both cases since it is assumed to be spherically symmetric. The surface brightness profile of the edge-on galaxy is two magnitudes brighter than the face-on counterpart because of the LOS integration. This LOS integration brings the brightness of the galaxy at the truncation radius well above the stellar halo contribution, allowing us to detect it. In the face-on case, the truncation in the stellar disc is completely outshone by the brightness of the stellar halo.}
\label{img:quali}
\end{figure}

We present in Fig.\ref{img:quali}, qualitatively, the main idea of this Letter: truncations in the surface brightness profile of low inclined spiral galaxies are outshone by the light of the  stellar halo component. 

The explanation for the lack of observational evidence of truncations in the radial light distribution of low-inclined galaxies is based on the different response that the stellar disk and the stellar halo have to the LOS integration. Whereas the surface brightness profile of a highly symmetric stellar halo is almost independent of orientation with respect to the observer, we expect from the model that the stellar disc is $\sim 2 ~ \mathrm{mag~arcsec}^{-2}$ dimmer in a face-on than in an edge-on projection. It is the absence of contrast between the stellar disc and the stellar halo which prevents us from detecting truncations in the surface brightness profiles of low-inclined spiral galaxies.

However, our model does not necessary imply that truncations cannot be studied in face-on galaxies. The contribution of the stellar halo to the total light of a galaxy depends on the galaxy mass \citep[see Fig.10 of][]{cou11}. The lower the mass of the galaxy, the weaker the contribution of the stellar halo to the total light is. Low-mass spirals are then the most suitable candidates to distinguish between truncations and stellar haloes in face-on disc galaxies. Another option to detect truncations in low-inclined systems could be through the decomposition of the surface brightness profiles in separate components: bulge, thin and thick discs, stellar halo... However, since we do not yet have a complete characterisation of the radial light distribution of stellar haloes, the subtraction of such component from the observed surface brightness distribution of a galaxy may lead to misleading conclusions.

The properties of the underlying stellar population could be also a tool to study the mechanism behind the formation of breaks and truncations in face-on spiral galaxies, without the inherent problems associated to the edge-on projection (e.g., dust attenuation and the mixing of stellar properties because of the LOS integration). The stellar disc and the stellar halo are expected to have very different kinematical properties and that can be used to spectroscopically differentiate the two components. Also, since truncations are thought to be related to the total angular momentum of the protogalactic cloud \cite{van87,van88}, a detailed study of the kinematics around the truncation radii might bring precious information about the angular momentum distribution when galaxies were formed. In addition, in terms of $\alpha$-element overabundance, age and metallicity, stellar discs and stellar haloes might show significant differences \citep{rs12}, with the stars in the stellar halo probably being older, more metal-poor and more $\alpha$-enhanced than the stars in the stellar disc. Deep enough resolved spectroscopic surveys such as CALIFA \citep{califa} and MaNGA\footnote{\textit{http://www.sdss3.org/future/manga.php}} might help to separately study truncations and stellar haloes. Finally the next generation of space (GAIA, James Webb Space Telescope) and ground based telescopes (European Extremely Large Telescope) will allow us to perform resolved stellar population analysis in the Milky Way and beyond the Local Group, opening the door to very detailed star counting studies of breaks and truncation up to the distance of the Virgo Cluster.

\vspace{0.2in}

\small{
\footnotesize{\textit{\textbf{Acknowledgements}}} IMN would like to thank Marja Seidel for her stimulating comments on this letter.

This work has been supported by the Programa Nacional de Astronom\'ia y Astrof\'isica of the Spanish Ministry of Science and Innovation under grant AYA2010-21322-C03-02. We acknowledge financial support to the DAGAL network from the People Programme (Marie Curie Actions) of the European Union's Seventh Framework Programme FP7/2007-2013/ under REA grant agreement number PITN-GA-2011-289313. 

Funding for SDSS-III has been provided by the Alfred P. Sloan Foundation, the Participating Institutions, the National Science Foundation, and the U.S. Department of Energy Office of Science. The SDSS-III web site is http://www.sdss3.org/. SDSS-III is managed by the Astrophysical Research Consortium for the Participating Institutions of the SDSS-III Collaboration including the University of Arizona, the Brazilian Participation Group, Brookhaven National Laboratory, University of Cambridge, Carnegie Mellon University, University of Florida, the French Participation Group, the German Participation Group, Harvard University, the Instituto de Astrof\'isica de Canarias, the Michigan State/Notre Dame/JINA Participation Group, Johns Hopkins University, Lawrence Berkeley National Laboratory, Max Planck Institute for Astrophysics, Max Planck Institute for Extraterrestrial Physics, New Mexico State University, New York University, Ohio State University, Pennsylvania State University, University of Portsmouth, 
Princeton University, the Spanish Participation Group, University of Tokyo, University of Utah, Vanderbilt University, University of Virginia, University of Washington, and Yale University. 

This research has made use of NASA's Astrophysics Data System and of the HyperLeda database.

}

\bibliographystyle{mn2e}
\bibliography{PhDsample}

\label{lastpage}

\end{document}